\documentclass[12pt]{article}
\usepackage{amsmath,amssymb}
\usepackage{amsthm}
\usepackage{mathrsfs}
\numberwithin{equation}{section}
\usepackage{cite}

\newcommand{\na}{\nabla}

\renewcommand{\b}{\bar}
\renewcommand{\d}{\dot}

\newcommand{\df}{\dfrac}

\newcommand{\ul}{\underline}
\newcommand{\der}{\partial}

\renewcommand{\(}{\left(}
\renewcommand{\)}{\right)}

\newcommand{\wed}{\wedge}
\newcommand{\bmx}{\left(\begin{matrix}}
\newcommand{\emx}{\end{matrix}\right)}

\newdimen\Tdim
\newdimen\Ddim
\def\Tspan#1{{\setbox0=\hbox{$#1$}%
\Tdim\ht0\advance\Tdim\dp0\advance\Tdim.7ex\Ddim\dp0\advance\Ddim.4ex\rule[-\Ddim]{0pt}{\Tdim}\box0}}

\usepackage[top=2.828cm,bottom=2.828cm, left=2.5cm,right=2.5cm]{geometry}
\newcommand{\parag}[1]{
\vspace{1ex}
\par
\noindent
\textit{#1}
\vspace{1ex}
\par
}
\begin{document}
\begin{titlepage}
\hfill 
\vspace{1em}
\def\thefootnote{\fnsymbol{footnote}}%
   \def\@makefnmark{\hbox
       to\z@{$\m@th^{\@thefnmark}$\hss}}%
 \vspace{2em}
 \begin{center}%
  {\Large \bf 
Abelian tensor hierarchy and Chern--Simons actions
\\[1ex] 
in 4D ${\cal N}=1$ conformal supergravity
  }%
  \par
 \vspace{1.5em} 
  {\large
    Ryo Yokokura\footnote{email: ryokokur@rk.phys.keio.ac.jp}
   \par
   }%
  \vspace{1em} 
{\small\it Department of Physics,
Keio  University,
Yokohama 223-8522, Japan
}
   {\large 
    }%
 \end{center}%
 \par
\vspace{1.5em}%
 \setcounter{footnote}{0}%
\def\thefootnote{\arabic{footnote}}%
   \def\@makefnmark{\hbox
       to\z@{$\m@th^{\@thefnmark}$\hss}}%
\begin{abstract}
We consider Chern--Simons actions
 of Abelian tensor hierarchy of $p$-form gauge fields
in four-dimensional ${\cal N}=1$ supergravity.
Using conformal superspace formalism, 
we solve 
the constraints on the field strengths of the $p$-form gauge 
superfields in the presence of the tensor hierarchy.
The solutions
 are expressed by the prepotentials
of the $p$-form gauge superfields.
We show
the internal and superconformal transformation laws of 
the prepotentials.
The descent formalism for the 
Chern--Simons actions is exhibited.
\end{abstract}
\end{titlepage}
\section{Introduction}
\label{sec:intro}
The superstring theory is a candidate for the unified theory 
of the fundamental interactions including quantum gravity.
There are strings and branes in the superstring theory.
Our universe can be described by the strings and branes
in a unified way at the low energy limit.
The stability of the branes is preserved by 
supersymmetry (SUSY) and conserved Ramond--Ramond charges.
We can avoid unstable tachyons by SUSY.
Further, the conserved charges guarantee
 the number of branes.
Antisymmetric tensor ($p$-form) 
 gauge fields are coupled to the conserved charges of the branes.

One of the most important issues in the superstring theory
 is to construct realistic four-dimensional (4D) effective theories.
Since the superstring theory is a ten-dimensional theory,
4D effective theories are obtained by compactifying
extra six dimensions.
4D ${\cal N}=1 $ supergravity (SUGRA) is a 
candidate for the effective theories.
This theory consists of chiral fermions as well as gravity.
Thus, we can embed the standard model particles into
the theory. 
Further, the stability of the theory is ensured by SUSY.

Thus, it is important
to consider $p$-form gauge fields in 4D ${\cal N}=1$ 
SUGRA~\cite{{Siegel:1979ai},{bib:G},{bib:SS},{Muller:1985vga},{Cecotti:1987qr},{Binetruy:1996xw},{Ovrut:1997ur},{Louis:2004xi},{D'Auria:2004sy},{Kuzenko:2004tn},{Louis:2007nd}}.
In particular, we study the $p$-form gauge fields which can be
regarded as dimensionally reduced ones from higher dimensions.
Such 
 $p$-form gauge fields differ from those of simply defined in 4D.
Because the original gauge transformation laws of
the $p$-form gauge fields are given in higher dimensions,
the gauge transformations of the 
$p$-form gauge fields should contain
different rank forms in 4D.
The structure of the transformations is called a tensor 
hierarchy~\cite{{deWit:2005hv},{deWit:2008ta},{deWit:2008gc},{Hartong:2009az}}.

In 4D ${\cal N}=1$ global SUSY, Becker et al.~constructed such Abelian tensor hierarchy in superspace~\cite{bib:BBLR}.
They showed 
Chern--Simons (CS) actions%
\footnote{They also showed CS actions in the 
case of non-Abelian tensor hierarchy~\cite{{Becker:2016rku}}.}.
The CS actions are constructed by integrands
which are proportional to $p$-form gauge fields.
Since the different ranked tensors are related each other 
by the tensor hierarchy, 
each of the integrands is not independent.
The internal gauge invariance requires the 
relations between the integrands.
They showed 
the relations in a systematic manner, which is called descent formalism.
The descent formalism relates the integrands each other by derivatives.
The CS actions are important because they are 
related to the anomaly cancellation in 4D~
\cite{{Girardi:1986zn},{LopesCardoso:1991ifk},{Adamietz:1992dk}}.

In this paper, we embed the CS actions
 of Abelian tensor hierarchy obtained in Ref.~\cite{bib:BBLR}
into 4D ${\cal N}=1$ SUGRA.
We use 4D ${\cal N}=1$ conformal superspace 
formalism~\cite{bib:B1}.
This formalism has larger gauge symmetries than 
 superconformal tensor
 calculus \cite{{bib:KTVN},{bib:KT},{bib:TVN},{bib:FGVN},{bib:CFGVP},{bib:KUigc},{Kugo:1982cu},{bib:KU},{bib:KKLVP}}
and  Poincar\'e superspace formalism~\cite{{bib:WB},{bib:BG2}}.
Superconformal tensor calculus and Poincar\'e superspace
 formalism 
 are obtained from the conformal superspace formalism 
by using their correspondences~\cite{{bib:B1},{bib:KYY},{Kugo:2016lum}}.
The CS actions are constructed by
 the prepotentials of the $p$-form gauge superfields
in the presence of the tensor hierarchy%
\footnote{In this paper, we use the term ``prepotentials''
to refer to superfields which consist of the bosonic gauge fields
and field strengths as well as their superpartners.}.
The prepotentials are obtained by so-called covariant approach,
which are shown in our previous paper~\cite{bib:AHYY}.
In the covariant approach, we introduce $p$-form gauge 
superfields and their field strength superfields in 
the superspace.
The field strength superfields have some constraints,
since they have superfluous degrees of freedom.
The prepotentials are obtained as the solutions to the 
constraints.
The CS actions in 4D ${\cal N}=1$ SUGRA 
 would be useful to discuss 
the roles of the $p$-form gauge fields, 
e.g. in cosmology~\cite{{Kaloper:2008fb},{Kaloper:2011jz}}.

In the conformal superspace, the derivations of 
 the solutions to the constraints are mostly the same
as the case of the global SUSY in Ref.~\cite{bib:G}.
This is because the anti-commutation relations of
 the spinor derivatives are the same forms as those of 
global SUSY.
Moreover, the descent formalism of the CS actions is also the 
same form as global SUSY case~\cite{bib:BBLR}, 
since the relation between 
D- and F-term integrations 
in the conformal superspace are quite similar to the 
global SUSY case.

This paper is organized as follows.
In section \ref{sec:rev}, we briefly review 
the covariant approach to Abelian tensor hierarchy 
in 4D ${\cal N}=1$ conformal superspace.
The prepotentials of $p$-form gauge superfields
are obtained in section \ref{sec:pp}.
We show the internal gauge transformation laws of the prepotentials. 
Section \ref{sec:cs} is devoted to constructing the CS actions of the tensor hierarchy.
In particular, the descent formalism 
in the conformal superspace is discussed.
Finally, we conclude this paper in section \ref{sec:conc}.
Throughout this paper, we use 
the terms ``gauge superfields'', ``field strengths superfields'',
and ``gauge parameter superfields'' are 
simply written as ``gauge fields'', ``field strengths'',
and ``gauge parameters'', respectively.
\section{Review of the covariant approach}
\label{sec:rev}
We briefly review so-called covariant approach
to Abelian tensor hierarchy in 4D ${\cal N}=1$ 
conformal superspace discussed in Ref.~\cite{bib:AHYY}.
Covariant approach is an approach to constructing 
supersymmetric theories of $p$-form gauge fields
in superspace.

We use the notations
 and conventions of Ref.~\cite{bib:AHYY}
except the normalizations of the superfields $Y^{I_3}$ and 
$L^{I_2}$, which  are the same as $G^{S}$ and $H^{M}$
in Ref.~\cite{bib:BBLR}, respectively.

\subsection{Conformal superspace}
We firstly review conformal superspace formalism to construct 
SUGRA~\cite{bib:B1}.
Superspace is space which is spanned by the ordinary spacetime
 coordinates $x^m $ and 
the Grassmannian coordinates $(\theta^\mu,\b\theta_{\d\mu})$.
Here, the indices $m,n,...$ are used to refer to curved vector indices.
The indices $\mu,\nu,...$ and $\d\mu,\d\nu,...$ denote
curved undotted and dotted spinor indices, respectively.
In the superspace,
 SUSY transformations are understood as the translations to 
 Grassmannian coordinates.
Thus, we simply denote these coordinates at the same time:
 $z^M=(x^m,\theta^\mu,\b\theta_{\d\mu})$,
where we use Roman capital indices $M,N,...$ for both of 
curved vector and spinor indices.

Conformal superspace is superspace where the superconformal symmetry 
is introduced as a gauge symmetry.
The generators of the superconformal symmetry are
spacetime translations $P_a$, SUSY transformations 
$(Q_\alpha,\b{Q}^{\d\alpha})$,
 Lorentz transformations $M_{ab}$, 
dilatation $D$, chiral rotation $A$, conformal boosts $K_a$, 
and conformal SUSY transformations $(S_\alpha,\b{S}^{\d\alpha})$.
Here, Roman letters $a,b,...$ denote flat vector indices.
Greek letters $\alpha,\beta,...$ and $\d\alpha,\d\beta,...$
denote flat spinor indices.
All of the generators of the superconformal symmetry are 
denoted 
as $X_{\cal A}$, where we use  calligraphic indices 
${\cal A,B,...}$ to
refer to the generators of the superconformal symmetry.
In the conformal superspace, 
both of $P_a$ and $(Q_\alpha,\b{Q}^{\d\alpha})$ are
understood as the translations.
Thus, we simply express $P_a$ and
$(Q_\alpha,\b{Q}^{\d\alpha})$ at the same time:
$P_A:=(P_a,Q_\alpha,\b{Q}^{\d\alpha})$.
Here, capital Roman letters $A,B,...$ are used for
both of flat vector and spinor indices.
Similarly, 
we denote both of $K_a$ and $(S_\alpha,\b{S}^{\d\alpha})$
as $K_A:=(K_a,S_\alpha,\b{S}^{\d\alpha})$.
The (anti-)commutation relations of the generators are summarized 
in Ref.~\cite{bib:B1}.

The gauge fields of the superconformal symmetry are given by
\begin{equation}
 h_M{}^{\cal A} X_{\cal A}
:=E_M{}^A P_A
+\df{1}{2}\phi_M{}^{ab}M_{ba}
+B_M D
+A_M A
+f_M{}^A K_A,
\end{equation} 
where we assume that the vielbein $E_M{}^A$ is invertible,
and the inverse of the vielbein is denoted as $E_A{}^M$: 
$E_M{}^AE_A{}^N=\delta_M{}^N$ and $E_A{}^M E_M{}^B=\delta_A{}^B$.
Note that the gauge fields $h_M{}^{\cal A}$ are also expressed by 
differential forms on the conformal superspace as
\begin{equation}
 h^{\cal A}=dz^M h_M{}^{\cal A}.
\end{equation}
Here, we use the convention of Ref.~\cite{bib:WB} for the differential
forms.
The differential forms 
$dz^M=(dx^m,d\theta^\mu,d\b\theta_{\d\mu})$ are bases of 
the superforms on the conformal superspace.
The gauge transformation parameters are denoted as
\begin{equation}
 \xi^{\cal A}X_{\cal A}
=\xi^A P_A
+\df{1}{2}\xi(M)^{ab}M_{ba}
+\xi(D)D
+\xi(A)A
+\xi(K)^A K_A.
\end{equation}
We denote infinitesimal superconformal transformations 
 as $\delta_G( \xi^{\cal A}X_{\cal A})$.
The transformation laws of the gauge fields
$h_M{}^{\cal A}$ 
under the superconformal transformations other than $P_A$
are
given by
\begin{equation}
 \delta_G(\xi^{\cal B'}X_{\cal B'}) h_M{}^{\cal A}
=\der_M \xi^{\cal B'}\delta_{\cal B'}{}^{\cal A}
+h_M{}^{\cal C}\xi^{\cal B'}f_{\cal B'C}{}^{\cal A}.
\end{equation}
Here, primed calligraphic indices 
${\cal A',B',...}$ are used to refer to 
the generators of the superconformal symmetry other than $P_A$:
$X_{\cal A'}=(M_{ab},D,A,K_A)$. 
The coefficients  $f_{\cal CB}{}^{\cal A}$ are 
the structure constants of the superconformal symmetry: 
$[X_{\cal C},X_{\cal B}]=-f_{\cal CB}{}^{\cal A} X_{\cal A}$,
 where we use the convention of ``implicit grading''~\cite{bib:B1}.

We define SUSY transformations and spacetime translations in 
the conformal superspace.
In the conformal superspace, SUSY transformations are regarded as 
translations to the Grassmannian coordinates.
Using field-independent parameters $\xi^A$,
we relate infinitesimal 
$P_A$-transformations $\delta_G(\xi^AP_A)$
to the general coordinate transformations 
$\delta_{\text{GC}}(\xi^M)$  as
\begin{equation}
\delta_G(\xi^AP_A) = 
\delta_{\text{GC}}(\xi^M)
-\delta_G(\xi^M{h_M}^{{\cal B}'}X_{{\cal B}'}).
\end{equation}
Here, the parameters $\xi^M$ are related to $\xi^A$ 
 as $\xi^M=\xi^A E_A{}^M$.
The actions of $P_A$-transformations on a superfield without curved
indices $\Phi$ define superconformally  covariant
derivatives $\na_A$:
\begin{equation}
 \delta_G(\xi^A P_A) \Phi
= \xi^A \na_A \Phi
=\xi^A E_A{}^M(\der_M-h_M{}^{\cal B'}X_{\cal B'}) \Phi.
\end{equation} 
\subsection{Covariant approach to Abelian tensor hierarchy}
Next, we introduce $p$-form gauge fields in the conformal superspace,
where $p$ runs over $p=-1,0,1,2,3,4$.
We assume that $(-1)$-forms are zero as 
in ordinary differential geometry.
The $p$-form gauge fields are denoted as 
\begin{equation}
 C^{I_p}_{[p]}
=\df{1}{p!}dz^{M_1}\wed \cdots \wed dz^{M_p} C_{M_p...M_1}^{I_p}
=\df{1}{p!}E^{A_1}\wed \cdots \wed E^{A_p} C_{A_p...A_1}^{I_p}.
\end{equation}
Here, $I_p$ are indices of internal degrees of freedom,
which run over $I_p=1,..., \dim V_p$.
The ranks of the differential forms are represented as $[p]$.
The $X_{\cal A'}$-transformations of the $p$-form gauge fields are
defined as
\begin{equation}
 \delta_G(\xi^{\cal A'}X_{\cal A'}) C^{I_p}_{M_p...M_1}=0.
\end{equation}
Thus, the $X_{\cal A'}$-transformations of $C_{A_p...A_1}^{I_p}$ are 
given by the $X_{\cal A'}$-transformations of vielbein $E_M{}^A$:
\begin{equation}
\begin{split}
 \delta_G(\xi^{\cal A'}X_{\cal A'}) C^{I_p}_{A_p...A_1}&=
- E_{A_p}{}^N (\delta_G(\xi^{\cal A'}X_{\cal A'})E_N{}^B) 
C^{I_p}_{B A_{p-1}...A_1}\\
&\quad 
-\cdots
-
E_{A_1}{}^N (\delta_G(\xi^{\cal A'}X_{\cal A'})E_N{}^B) 
C^{I_p}_{A_p ...A_2 B}.
\end{split}
\label{eq:scfgf}
\end{equation}
The explicit transformation of the vielbein is 
summarized in Ref.~\cite{bib:AHYY}.
The infinitesimal internal gauge transformations 
$\delta_T(\Lambda)$ of the $p$-form gauge fields are given by
\begin{equation}
 \delta_T(\Lambda) C^{I_p}_{[p]}
= d\Lambda_{[p-1]}^{I_p} +(q^{(p)}\cdot \Lambda_{[p]})^{I_p}.
\label{eq:gt}
\end{equation}
Here, $d$ denotes the exterior derivative in the conformal superspace,
and $\Lambda$ is the set of the real gauge parameter superforms:
$\Lambda=(\Lambda^{I_1}_{[0]},...,\Lambda^{I_4}_{[3]})$.
We assume that $\Lambda^{I_p}_{M_{p-1}...M_1}$ are field independent 
parameters.
Note that 
$\Lambda^{I_p}_{A_{p-1}...A_1}
=E_{A_{p-1}}{}^{M_{p-1}}\cdots E_{A_1}{}^{M_1}
\Lambda^{I_p}_{M_{p-1}...M_1}$ are field dependent parameters.
Ordinary Abelian gauge transformations are 
expressed by the first term in Eq.~\eqref{eq:gt}.
Shifts of the gauge fields are represented by the second term due to the tensor hierarchy.
$q^{(p)}$ are real linear maps from the vector space $V_{p+1}$ 
to the vector space $V_p$.
The expressions $(q^{(p)}\cdot \Lambda_{[p]})^{I_p}$ mean
$(q^{(p)})^{I_p}_{I_{p+1}} \Lambda_{[p]}^{I_{p+1}}$.
Note that $q^{(p)}$ can be understood as 
the exterior derivative
on the extra dimensions~\cite{bib:BBLR}.

The $P_A$-transformations are 
redefined with respect to the internal gauge transformations
in the presence of the tensor hierarchy.
The redefinitions are given by 
\begin{equation}
\delta_G(\xi^AP_A) = 
\delta_{\text{GC}}(\xi^M)
-\delta_G(\xi^M{h_M}^{{\cal B}'}X_{{\cal B}'})
-\delta_T(\Lambda(\xi)).
\end{equation}
Here, $ \Lambda(\xi)$ is defined by
\begin{equation}
 \Lambda(\xi)
=(\iota_\xi C^{I_1}_{[1]},...,\iota_\xi C^{I_4}_{[4]}),
\end{equation}
 and $\iota_\xi$ is a interior product
\begin{equation}
 \iota_\xi C^{I_p}_{[p]}
=\df{1}{(p-1)!}
dz^{M_1}\wed \cdots \wed dz^{M_{p-1}}
\xi^{M_p}C_{M_p ... M_1}^{I_p}.
\end{equation}

In the presence of the tensor hierarchy, 
the field strengths of the $p$-form gauge fields are given by
using the exterior derivative and $q$'s.
The definitions of the field strengths of the $p$-form gauge 
fields are given as follows:
\begin{equation}
 F^{I_p}_{[p+1]}
=dC^{I_p}_{[p]}- (q^{(p)}\cdot C_{[p+1]})^{I_p}.
\label{eq:fs}
\end{equation}
The field strengths are transformed under the internal gauge
transformations as
\begin{equation}
 \delta_T(\Lambda) F_{[p+1]}^{I_p}
=-(q^{(p)}\cdot q^{(p+1)} \cdot \Lambda_{[p+1]})^{I_p}.
\end{equation}
The invariances of the field strengths under the internal 
transformations require conditions on the $q$'s as
\begin{equation}
 q^{(p)}\cdot q^{(p+1)}=0.
\end{equation}
The covariant derivatives on the field strengths with 
Lorentz indices are given by
\begin{equation}
 \na_B F^{I_p}_{A_{p+1}...A_1}
=E_B{}^M (\der_M-{h_M}^{{\cal A}'}X_{{\cal A}'})
F^{I_p}_{A_{p+1}...A_1}.
\label{eq:nafs}
\end{equation}
Note that 
the covariant derivatives $\na_B$ on the field strengths 
$F^{I_p}_{A_{p+1}...A_1}$
are  superconformally covariant 
and internally invariant derivatives because
$F^{I_p}_{A_{p+1}...A_1}$ are 
invariant under the internal gauge transformations.
The Bianchi identities for the field strengths are given by
\begin{equation}
0=dF^{I_p}_{[p+1]}+ (q^{(p)} \cdot F_{[p+2]})^{I_p}.
\end{equation}
We summarize the explicit forms of the gauge fields,
field strengths and Bianchi identities in table \ref{tab:gfb}.
\begin{table}[t]
\[
\begin{array}{c|c|c|c}
\hline\hline
\text{form}&\text{gauge field}&\text{field strength}
 &\text{Bianchi identity}
\\
\hline
\text{4-form}&\Tspan{U^{I_4}}
&G^{I_4}=dU^{I_4}=0&
-
\\
\text{3-form}&
\Tspan{C^{I_3}}
&\Sigma^{I_3}=dC^{I_3}-(q^{(3)}\cdot U)^{I_3}
&d\Sigma^{I_3}=0
\\
\text{2-form}&
B^{I_2}&H^{I_2}=dB^{I_2}-(q^{(2)}\cdot C)^{I_2}
&dH=-(q^{(2)}\cdot \Sigma)^{I_2}
\\
\text{1-form}&A^{I_1}&F^{I_1}=dA^{I_1}-(q^{(1)}\cdot B)^{I_1}
& dF^{I_1}=-(q^{(1)}\cdot H)^{I_1}
\\
\text{0-form}&f^{I_0}&g^{I_0}=df^{I_0}-(q^{(0)}\cdot A)^{I_0}
& dg^{I_0}=-(q^{(0)}\cdot F)^{I_0}
\\
\text{$-1$-form}&0&\omega^{I_{-1}} = -(q^{(-1)}\cdot f)^{I_{-1}}
& d\omega^{I_{-1}}=-(q^{(-1)}\cdot g)^{I_{-1}}
\\
\hline\hline
 \end{array}  
\]
 \caption{
The $p$-forms, their corresponding field strengths and Bianchi
 identities.
We impose  that 
the field strengths of the 4-form gauge fields are zero
 as in table \ref{tab:constr}.}
\label{tab:gfb}
\end{table}

We impose some constraints on the field strengths 
to eliminate degrees of freedoms because
 there are superfluous degrees of freedoms 
in the field strengths in the superspace.
The constraints are given as in table \ref{tab:constr},
 which are the same forms 
as the cases that the tensor hierarchy does not exist 
\cite{{bib:G},{bib:SS},{bib:BG2}}.
In this table, the indices $\ul\alpha, \ul\beta,...$ denote both 
undotted and dotted spinor indices: $\ul\alpha=(\alpha,\d\alpha)$.
\begin{table}[t]
\[
\begin{array}{c|c}
\hline\hline
\text{form}&\text{constraints}
\\
\hline
\text{4-form}&
\Tspan{ 
G^{I_4}_{EDCBA}=0
}
\\
\hline
\text{3-form}&
\Tspan{
  \Sigma^{I_3}_{\ul\delta\, \ul\gamma\,\ul\beta A}
=\Sigma^{I_3}_{\delta\d\gamma ba}=0
}
\\
\hline
\text{2-form}&
\Tspan{
 H^{I_2}_{\ul\gamma\,\ul\beta\,\ul\alpha}
=H^{I_2}_{\gamma\beta a}=H^{I_2}_{\d\gamma\d\beta a}=0,
\quad
H^{I_2}_{\gamma\d\beta a}
=i(\sigma_a)_{\gamma\d\beta}L^{I_2}
}
\\
\hline
\text{1-form}
&
\Tspan{
 F^{I_1}_{\ul\alpha\,\ul\beta}=0
}
\\
\hline
\text{0-form}& 
\Tspan{
g^{I_0}_\alpha =i \na_\alpha \Psi^{I_0}, 
\quad 
g^{I_0}_{\d\beta}=-i\b\na_{\d\beta} \Psi^{I_0},
\quad 
K_A \Psi^{I_0}=0
}
\\
\hline\hline
 \end{array} 
\]
\caption{The constraints on the field strengths.
}
\label{tab:constr} 
\end{table}
Note that the constraints are covariant 
under both superconformal and internal gauge transformations.

We solve the Bianchi identities under the constraints.
As a result, the field strengths 
are expressed by the irreducible superfields.
The irreducible superfields of the 2- and 0-form gauge fields are
$L^{I_2}$ and $\Psi^{I_0}$  in table \ref{tab:constr}.
We find the irreducible superfields of 3- and 1-form gauge 
fields $Y^{I_3}$ and $W_{\ul\alpha}^{I_1}$ as follows, 
respectively:
\begin{equation}
\Sigma^{I_3\d\delta\d\gamma}{}_{ba}
=
4(\b\sigma_{ba}\epsilon)^{\d\delta\d\gamma} Y^{I_3},
\quad
\Sigma^{I_3}_{\delta\gamma ba}
=4(\sigma_{ba}\epsilon)_{\delta\gamma} \b{Y}^{I_3}.
\label{eq:3sf}
\end{equation}
\begin{equation}
 F_{\d\beta,\alpha\d\alpha}^{I_1}
=-2\epsilon_{\d\beta\d\alpha} W_{\alpha}^{I_1},
\quad
F^{I_1}_{\beta,\alpha\d\alpha}
=-2\epsilon_{\beta\alpha}
\b{W}^{I_1}_{\d\alpha}.
\label{eq:1sf}
\end{equation}
Note that the Weyl weights $\Delta$ and chiral
weights $w$ of the irreducible superfields
are as follows:
\begin{equation}
\begin{split}
 Y^{I_3}&: (\Delta,w)=(3,2),\\
 L^{I_2}&: (\Delta,w)=(2,0),\\
 W_\alpha^{I_1}&: (\Delta,w)=(3/2,1),\\
 \Psi^{I_0}&: (\Delta,w)=(0,0).
\end{split}
\label{eq:cwsf}
\end{equation}
Here, Weyl and chiral weights of 
a superfield $\Phi$ are given by
\begin{equation}
 D \Phi=\Delta \Phi,\quad A \Phi= iw\Phi.
\end{equation}
Hereafter, we use the term ``conformal weights'' to refer to
`` Weyl and chiral weights''.

The  tensor hierarchy deforms the properties of 
the irreducible superfields such as 
the linearity conditions for $L^{I_2}$ and reality conditions for
$W_{\ul\alpha}^{I_1}$:
\begin{equation}
\begin{split}
& -\df{1}{4} \b\na^2 L^{I_2}= -(q^{(2)}\cdot Y)^{I_2},
\quad  -\df{1}{4} \na^2 L^{I_2}= -(q^{(2)}\cdot \b{Y})^{I_2},
\\
& \df{1}{2i}(\na^\alpha
 W_{\alpha}^{I_1}-\b\na_{\d\alpha}\b{W}^{I_1\d\alpha})
=-(q^{(1)}\cdot L)^{I_1},
\\
&
-\df{1}{4}\b\na^2 \na_\alpha \Psi^{I_0}
=- (q^{(0)}\cdot W_\alpha)^{I_0},
\quad
-\df{1}{4}\na^2 \b\na_{\d\alpha} \Psi^{I_0}
=- (q^{(0)}\cdot \b{W}_{\d\alpha})^{I_0}.
\end{split}
\end{equation}
Note that the derivatives $\na_A$ on the superfields $Y^{I_3}$,
$L^{I_2}$,
$W_{\ul\alpha}^{I_1}$, and $\Psi^{I_0}$ are 
superconformally covariant 
and internally invariant derivatives because of the 
properties in Eq.~\eqref{eq:nafs}.
\section{Prepotentials}
\label{sec:pp}
In this section, we construct the prepotentials of 
the $p$-form gauge fields in the presence of the tensor hierarchy.
The prepotentials and their gauge transformation laws 
are needed to construct CS actions.
The prepotentials are obtained by solving
the constraints on the field strengths
 in certain gauge-fixing conditions.
The relations between the prepotentials and 
the irreducible superfields  are also 
obtained by the relations of the gauge fields and field strengths
in Eq.~\eqref{eq:fs}.
The gauge transformations of the prepotentials are determined 
by the gauge transformations which leave the gauge-fixing conditions 
invariant.
\subsection{Gauge-fixing conditions for the $p$-form  gauge fields}
We solve the constraints on the field strengths.
Since the constraints in table \ref{tab:constr} 
are gauge covariant,
we solve the constraints under the gauge-fixing conditions
where
some components of the gauge fields are gauged away
by using the definitions of the field strengths 
\begin{equation}
\begin{split}
F_{[p+1]}^{I_p}
&=
\df{1}{p!}E^{A_1}\wed \cdots \wed E^{A_{p}} \wed E^B 
\na_B C^{I_p}_{A_p...A_1}
\\
&\quad
+\df{1}{p!2!}E^{A_1}\wed \cdots E^{A_{p-2}}
\wed E^{B}\wed E^C
T_{CB}{}^{A_p}C^{I_p}_{A_p...A_1}\\
&\quad
+\df{1}{(p+1)!}E^{A_1}\wed \cdots \wed E^{A_{p+1}} 
(q^{(p)}\cdot C_{A_{p+1}...A_1})^{I_p},
\end{split}
\label{eq:covfs}
\end{equation}
and the internal gauge transformation laws of the gauge fields
\begin{equation}
\begin{split}
\delta_T(\Lambda)C_{[p]}^{I_p}
&=
\df{1}{(p-1)!}E^{A_1}\wed \cdots \wed E^{A_{p-1}} \wed E^B 
\na_B \Lambda^{I_p}_{A_{p-1}...A_1}
\\
&\quad
+\df{1}{(p-1)!2!}E^{A_1}\wed \cdots E^{A_{p-2}}
\wed E^{B}\wed E^C
T_{CB}{}^{A_{p-1}}\Lambda^{I_p}_{A_{p-1}...A_1}\\
&\quad
+\df{1}{p!}E^{A_1}\wed \cdots \wed E^{A_p} 
(q^{(p)}\cdot \Lambda_{A_p...A_1})^{I_p}.
\end{split}
\label{eq:covgt}
\end{equation}
Here, $\na_A$ are covariant 
with respect to only the superconformal symmetry,
and $T_{CB}{}^A$ are the coefficients of torsion 2-form 
defined by 
\begin{equation}
 T^A
=\df{1}{2}E^B \wed E^C T_{CB}{}^A
=dE^A-E^C\wed h^{{\cal B}'}f_{{\cal B}'C}{}^A.
\end{equation}
The gauge-fixing conditions are the same form as the case 
of global SUSY \cite{bib:G} because of the following
 three reasons.
First, the constraints on the following components of 
the torsion are the same as those of global SUSY 
(see Ref.~\cite{bib:B1}):
\begin{equation}
 T_{\gamma\beta}{}^A=0,
\quad
 T_{\d\gamma\d\beta}{}^A=0,
\quad
 T_{\gamma\d\beta}{}^a=2i(\sigma^a)_{\gamma\d\beta},
\quad
T_{\gamma\d\beta}{}^{\ul\alpha}=0,
\quad
T_{\ul{\gamma} b}{}^A=0,
\quad
T_{cb}{}^a=0.
\end{equation}
Second, as announced in section \ref{sec:intro}, 
the anti-commutation relations and of the superconformally
 covariant spinor derivatives are the same form as those of 
global SUSY case:
\begin{equation}
 \{\na_\alpha,\na_\beta\}=0,
\quad
 \{\b\na_{\d\alpha},\b\na_{\d\beta}\}=0,
\quad
 \{\na_{\alpha},\b\na_{\d\beta}\}=-2i \na_{\alpha\d\beta}.
\end{equation}

Third, 
if we impose the gauge-fixing 
conditions and solve the constraints  
in order of 4-, 3-, 2, and 1-form,
the gauge-fixing conditions are not deformed  
from the case of the absence of 
the tensor hierarchy in Ref.~\cite{bib:G}.
For example, we discuss the gauge-fixing conditions for $C^{I_3}_{\gamma\beta\alpha}$.
Since the field strengths of the 4-form gauge fields 
are the same as the case of the absence of the tensor hierarchy,
we fix some of the 4-form gauge fields, e.g., 
$U^{I_4}_{\delta\gamma\beta\alpha}=0$.
Under the gauge-fixing conditions
$U^{I_4}_{\delta\gamma\beta\alpha}=0$,
the field strengths of the 3-form gauge fields
 $\Sigma^{I_3}_{\delta\gamma\beta\alpha}$ are written as
$ \Sigma^{I_3}_{\delta\gamma\beta\alpha}
= \na_{\delta}C^{I_3}_{\gamma\beta\alpha}
+\na_{\gamma}C^{I_3}_{\delta\beta\alpha}
+\na_{\beta}C^{I_3}_{\gamma\delta\alpha}
+\na_{\alpha}C^{I_3}_{\delta\gamma\beta}$.
We find that the terms 
$(q^{(3)}\cdot U_{\delta\gamma\beta\alpha})^{I_3}$
do not appear in the field strengths 
$\Sigma^{I_3}_{\delta\gamma\beta\alpha}$ in this gauge.
Thus, we impose the same gauge-fixing conditions 
as the case of the absence of the tensor hierarchy:
$C^{I_3}_{\gamma\beta\alpha}=0$,
which are derived from the constraints
$ \Sigma^{I_3}_{\delta\gamma\beta\alpha}=0$.

Therefore, the gauge-fixing conditions are 
the same as the case that the tensor hierarchy does not exist.
The explicit forms are summarized in table \ref{tab:gf}.
In this table, $X^{I_3}$ and 
$V^{I_1}$ are real superfields, which are the
prepotentials of the 3- and 1-form gauge fields, respectively.
\begin{table}[t]
\[
\begin{array}{c|c}
\hline\hline
\text{form}&\text{conditions on the gauge fields}
\\
\hline
\text{4-form}&
\Tspan{ 
 U^{I_4}_{\ul\delta\, \ul\gamma\, \ul\beta \, A}=
U^{I_4}_{\delta\d\gamma ba}=0
}
\\
\hline
\text{3-form}&
\Tspan{
 C^{I_3}_{\ul\gamma\,\ul\beta\, \ul\alpha}=
C^{I_3}_{\gamma\beta a}=
C^{I_3}_{\d\gamma\d\beta a}=0,
\quad
C^{I_3}_{\gamma \d\beta a}=i(\sigma_a)_{\gamma\d\beta} X^{I_3}
}
\\
\hline
\text{2-form}&
\Tspan{
 B^{I_2}_{\ul\beta\, \ul\alpha}=0
}
\\
\hline
\text{1-form}
&
\Tspan{
 A^{I_1}_\alpha = i\na_\alpha V^{I_1},
\quad 
A^{I_1}_{\d\alpha}=-i \b\na_{\d\alpha} V^{I_1}
}
\\
\hline\hline
 \end{array} 
\]
\caption{The gauge-fixing conditions on the gauge fields.
The gauge-fixing conditions are imposed in the order of 4-, 3-, 2- and
 1-form guage fields.
}
\label{tab:gf} 
\end{table}
\subsection{Prepotentials: The solutions to the constraints}
In this subsection,
 we show the prepotentials for the $p$-form gauge fields.
Under the gauge-fixing conditions 
and the constraints on the field strengths, 
the gauge fields are expressed in terms of the prepotentials.
We remark that the gauge-fixing conditions 
of $p$-form gauge fields are the same form 
as the constraints on the field strengths of $(p-1)$-form 
gauge fields.
Thus, we solve the constraints by the same procedure 
as the Bianchi identities for the field 
strengths~\cite{bib:AHYY}.
The conformal weights of the prepotentials are also determined
by using Eq.~\eqref{eq:scfgf}.
We exhibit the expressions of the gauge fields in terms of the
prepotentials as follows.
\parag{The 4-form gauge fields}
The solutions to the gauge-fixing conditions and constraints 
for the field strengths are the same as 
the case of the absence of the tensor hierarchy.
The prepotentials of the 4-form gauge fields are given as the 
2-spinor/2-vector components:
\begin{equation}
U^{I_4\d\delta\d\gamma}{}_{ ba}
=4(\b\sigma_{ba}\epsilon)^{\d\delta\d\gamma}
\Gamma^{I_4},
\quad
U^{I_4}_{\delta\gamma ba}
=4(\sigma_{ba}\epsilon)_{\delta\gamma}
\b\Gamma^{I_4}.
\label{eq:4pp}
\end{equation}
The prepotentials $\Gamma^{I_4}$ are primary superfields with  
conformal weights $(\Delta,w)=(3,2)$, which are derived from the 
superconformal transformation laws of 
$U^{I_4}_{\d\delta\d\gamma ba}$ in Eq.~\eqref{eq:scfgf}.
The prepotential $\Gamma^{I_4}$ and $\b\Gamma^{I_4}$ 
are chiral and anti-chiral superfields, respectively:
\begin{equation}
 \b\na_{\d\alpha}\Gamma^{I_4}=0,
\quad
\na_\alpha \b\Gamma^{I_4}=0.
\end{equation}
The other components the 4-form gauge fields 
are expressed in terms of the prepotentials 
\begin{equation}
 U^{I_4\d\delta}{}_{ cba}
=+\df{1}{2}(\b\sigma^d)^{\d\delta \delta}\epsilon_{dcba}
\na_\delta\Gamma^{I_4},
\quad
 U^{I_4}_{\delta cba}
=-\df{1}{2}(\sigma^d)_{\delta\d\delta}\epsilon_{dcba}
\b\na^{\d\delta}\b\Gamma^{I_4},
\end{equation}
\begin{equation}
 U^{I_4}_{dcba}
=\df{i}{8}\epsilon_{dcba}(\na^2\Gamma^{I_4} -\b\na^2\b\Gamma^{I_4}).
\end{equation}
\parag{The 3-form gauge fields}
We find the prepotentials of the 3-form gauge fields $X^{I_3}$ in the 
2-spinor/1-vector component,
where $X^{I_3}$ are real primary superfields with 
conformal weights $(\Delta,w)=(2,0)$.
The derivatives of the prepotentials give the other components 
of the gauge fields as 
\begin{equation}
 C^{I_3}_{\gamma ba}
=(\sigma_{ba})_\gamma{}^\delta\na_\delta X^{I_3},
\quad
 C^{I_3\d\gamma}{}_{ ba}
=
(\b\sigma_{ba})^{\d\gamma}{}_{\d\delta}\b\na^{\d\delta} X^{I_3}, 
\end{equation}
\begin{equation}
 C^{I_3}_{cba}
=\df{1}{8}\epsilon_{cbad}(\b\sigma^d)^{\d\delta\delta}
[\na_\delta,\b\na_{\d\delta}] X^{I_3}.
\end{equation}
\parag{The 2-form gauge fields}
The prepotentials of the 2-form gauge fields are  
primary superfields $\Sigma^{I_2}_\alpha$ and their conjugates 
$\b\Sigma^{I_2}_{\d\alpha}$.
The prepotentials are found in the spinor/vector components:
\begin{equation}
 B^{I_2}_{\beta,\alpha\d\alpha}
=-2 \epsilon_{\beta\alpha}\b\Sigma^{I_2}_{\d\alpha},
\quad
 B^{I_2}_{\d\beta,\alpha\d\alpha}
=-2\epsilon_{\d\beta\d\alpha}\Sigma^{I_2}_\alpha.
\label{eq:2pp}
\end{equation}
Here, $\Sigma^{I_2}_\alpha$ are primary superfields
 with conformal
 weights $(\Delta,w)=(3/2,1)$.
The prepotential $\Sigma^{I_2}_\alpha$ and $\b\Sigma^{I_2}_{\d\alpha}$
are chiral and anti-chiral superfields, respectively:
\begin{equation}
 \b\na_{\d\beta} \Sigma^{I_2}_\alpha=0, 
\quad 
 \na_\beta \b\Sigma^{I_2}_{\d\alpha}=0.
\end{equation}
The 2-vector components are as follows:
\begin{equation}
 B^{I_2}_{ba}
=\df{1}{2i}
\((\sigma_{ba})_\beta {}^\alpha 
\na^\beta \Sigma^{I_2 }_\alpha
-(\b\sigma_{ba})^{\d\beta}{}_{\d\alpha}
\b\na_{\d\beta}\b\Sigma^{I_2\d\alpha}
\).
\end{equation}
\parag{The 1-form gauge fields}
As in ordinary super QED case, the spinor components of
1-form gauge fields are given by 
real primary superfields $V^{I_1}$ in table \ref{tab:gf}.
The conformal weights of $V^{I_1}$ are $(\Delta,w)=(0,0)$.
The vector components are 
expressed by 
 \begin{equation}
 A^{I_1}_{\alpha\d\alpha}
=\df{1}{2}[\na_\alpha,\b\na_{\d\alpha}]V^{I_1}.
\end{equation}
We assume that $V^{I_1}$ are primary sueprfields: $K_A V^{I_1}=0$.
This assumption and conformal weights of $V^{I_1}$ 
are consistent with the $K_A$-invariances of 
$A_{\ul\alpha}^{I_1}$~\cite{bib:KU}.
\parag{The 0-form gauge fields}
The constraints on the field strengths of the 0-form are satisfied
if the gauge fields are real parts of chiral superfields 
$\Phi^{I_0}$, which are the prepotentials of 0-form gauge fields:
\begin{equation}
 f^{I_0}=\df{1}{2}(\Phi^{I_0}+\b\Phi^{I_0}).
\label{eq:0pp}
\end{equation}
Here, the conformal weights of $\Phi^{I_0}$ are 
$(\Delta,w)=(0,0)$,
and $\Phi^{I_0}$ are assumed to be primary superfields.
\parag{The relations between the prepotentials
 and the irreducible superfields}
We then find the relations 
 between the prepotentials and the irreducible superfields.
The relations are found as follows.
On the one hand, the irreducible superfields are given by 
the components of the field strengths 
$\Sigma^{I_3}_{\delta\gamma ba}$, 
$\Sigma^{I_3}_{\d\delta\d\gamma ba}$,
$H^{I_2}_{\gamma\d\beta a}$,
$F^{I_1}_{\ul\beta a}$,
and $g^{I_0}_a$.
On the other hand, the field strengths are expressed 
by the derivatives of the gauge fields in Eq.~\eqref{eq:fs}, which 
are now written in terms of the prepotentials.
In addition, the field strengths of $(-1)$-form gauge fields
 $\omega^{I_{-1}}$ are given by the 0-form gauge fields $f^{I_0}$
as in table~\ref{tab:gfb}: 
$\omega^{I_{-1}}=-(q^{(-1)}\cdot f)^{I_{-1}}$.
Since the 0-form gauge fields are expressed by the prepotential
 $\Phi^{I_0}$, the field strengths $\omega^{I_{-1}}$ are now
given by the real parts of chiral  superfields $J^{I_{-1}}=-(q^{(-1)}\cdot \Phi)^{I_{-1}}$:
\begin{equation}
\omega^{I_{-1}}
=\df{1}{2}(J^{I_{-1}}+\b{J}^{I_{-1}}).
\end{equation}

Thus, 
we find the relations by using the definitions of the
field strengths in terms of gauge fields \eqref{eq:fs},
the definitions of the superfields 
  in Eqs.~\eqref{eq:3sf}, \eqref{eq:1sf} and table~\ref{tab:constr}.

The results are summarized in table \ref{tab:fspp}.
Note that the irreducible 
superfields for $p$-form gauge fields are 
expressed by the prepotentials of $p$-  
and $(p+1)$-form gauge fields due to the tensor hierarchy.
\begin{table}[t]
\[
\begin{array}{c|c}
\hline\hline
\text{form}&\text{ prepotentials and irreducible superfields}
\\
\hline
\text{3-form}&
\Tspan{
Y^{I_3} = -\df{1}{4}\b\na^2 X^{I_3}  - (q^{(3)}\cdot \Gamma)^{I_3},
\quad
\b{Y}^{I_3} 
= -\df{1}{4}\na^2 X^{I_3}  - (q^{(3)}\cdot \b\Gamma)^{I_3}
}
\\
\hline
\text{2-form}&
\Tspan{
 L^{I_2}= \df{1}{2i}
(\na^\alpha \Sigma^{I_2}_\alpha 
- \b\na_{\d\alpha}\b\Sigma^{I_2 \d\alpha})
-(q^{(2)}\cdot X)^{I_2}
}
\\
\hline
\text{1-form}
&
\Tspan{
 W^{I_1}_\alpha
=-\df{1}{4}\b\na^2\na_\alpha V^{I_1} 
-(q^{(1)}\cdot \Sigma_\alpha)^{I_1},
\quad
\b{W}^{I_1}_{\d\alpha}
=-\df{1}{4}\na^2\b\na_{\d\alpha} V^{I_1}
-(q^{(1)}\cdot \b\Sigma_{\d\alpha})^{I_1}
}
\\
\hline
\text{0-form}& 
\Tspan{
 \Psi^{I_0}
=\df{1}{2i}(\Phi^{I_0}-\b\Phi^{I_0})-(q^{(0)}\cdot V)^{I_0}
}
\\
\hline
\text{$(-1)$-form}& 
\Tspan{
 J^{I_{-1}}=-(q^{(-1)}\cdot\Phi)^{I_{-1}}
}
\\
\hline\hline
 \end{array} 
\]
\caption{The relations between  the prepotentials
and the irreducible superfields.
}
\label{tab:fspp} 
\end{table}
\subsection{The gauge transformation laws of the prepotentials}

In this subsection, we show the transformation
 laws of the prepotentials.
The transformation laws are important when we construct 
CS actions. 
We have solved the gauge fields in terms of the prepotentials
under the set of the gauge-fixing conditions.
Although it seems that the gauge parameters are 
exhausted to fix the gauge fields,
there are 
remaining gauge parameters which preserve
 the gauge-fixing conditions 
in table \ref{tab:gf} invariant.
The remaining gauge transformation laws are determined 
 by 
the conditions 
for the gauge fields which are gauged away in table \ref{tab:gf}:
\begin{equation}
 0=\delta_T(\Lambda) C^{I_p}_{[p]}
=d\Lambda_{[p-1]}^{I_p}+ (q^{(p)}\cdot \Lambda_{[p]})^{I_p}.
\end{equation}

We denote the remaining parameters as 
$\Theta=(\Theta^{I_1},\Theta^{I_2},\Theta^{I_3}_{\ul\alpha},\Theta^{I_4})$.
We determine the properties of $\Theta$'s and 
the gauge transformation laws of the prepotentials
 as follows.
\parag{The 4-form gauge fields}
The gauge parameters are determined by the conditions so 
that the following gauge-fixing conditions are invariant:
\begin{equation}
 \delta_T(\Lambda) U^{I_4}_{\ul\delta\,\ul\gamma\, \ul\beta\, A}=0,
\quad
\delta_T(\Lambda) U^{I_4}_{\delta\d\gamma ba}=0.
\end{equation}
The gauge transformations which preserve the gauge-fixing conditions 
are given by
\begin{equation}
 \Lambda^{I_4}_{\ul\gamma \,\ul\beta\,\ul\alpha}=0,
\quad
\Lambda^{I_4}_{\gamma\beta a}=0,
\quad
\Lambda^{I_4}_{\d\gamma\d\beta a}=0,
\quad
 \Lambda^{I_4}_{\gamma \d\beta a}
=i (\sigma_a)_{\gamma\d\beta} \Theta^{I_4}.
\label{eq:4gp}
\end{equation}
Here, $\Theta^{I_4}$ are real superfields.
The prepotentials $\Gamma^{I_4}$ and $\b\Gamma^{I_4}$
are transformed by 
$\Theta^{I_4}$ as
\begin{equation}
 \delta_T(
\Lambda^{I_1},\Lambda^{I_2},\Lambda^{I_3},\Theta^{I_4})
\Gamma^{I_4}
=-\df{1}{4}\b\na^2 \Theta^{I_4},
\quad
 \delta_T(
\Lambda^{I_1},\Lambda^{I_2},\Lambda^{I_3},\Theta^{I_4})
\b\Gamma^{I_4}
=-\df{1}{4}\na^2 \Theta^{I_4},
\label{eq:4gtpp}
\end{equation}
which are determined by the gauge transformation laws of 
$U^{I_4}_{\d\delta\d\gamma ba}$ and 
$U^{I_4}_{\delta\gamma ba}$, respectively.

We can impose Wess--Zumino (WZ) gauge for the prepotentials 
$\Gamma^{I_4}$ by using $\Theta^{I_4}$ as follows:
\begin{equation}
 \Gamma^{I_4}|=0,
\quad
\na_\alpha \Gamma^{I_4}|=0,
\quad
 \b\na_{\d\alpha} \b\Gamma^{I_4}|=0,
\quad
(\na^2\Gamma^{I_4}+\b\na^2\b\Gamma^{I_4})|=0.
\label{eq:4wz}
\end{equation}
Here, the symbol of ``$|$'' means $\theta=\b\theta=0$ projection.
\parag{The 3-form gauge fields}
We impose that the following gauge-fixing conditions are 
invariant:
\begin{equation}
\begin{split}
& \delta_T( \Lambda^{I_1},\Lambda^{I_2},\Lambda^{I_3},\Theta^{I_4})
C^{I_3}_{\ul\gamma\,\ul\beta\,\ul\alpha}
=0,
\\
& \delta_T( \Lambda^{I_1},\Lambda^{I_2},\Lambda^{I_3},\Theta^{I_4})
C^{I_3}_{\gamma\beta a}
=0,
\quad
 \delta_T(\Lambda^{I_1},\Lambda^{I_2},\Lambda^{I_3},\Theta^{I_4})
C^{I_3}_{\d\gamma\d\beta a}
=0. 
\label{eq:3gi}
\end{split}
\end{equation}
The invariances are preserved 
by the conditions of the following gauge parameters:
\begin{equation}
 \Lambda^{I_3}_{\ul\beta\,\ul\alpha}=0.
\end{equation}
Note that the gauge parameters $\Theta^{I_4}$ 
do not change the gauge-fixing conditions in Eq.~\eqref{eq:3gi}
under the conditions for 
the gauge parameters in Eq.~\eqref{eq:4gp}
even if the tensor hierarchy exists.
Solving the constraints on the parameters,
we obtain that the remaining gauge parameters are
\begin{equation}
 \Lambda^{I_3}_{\d\beta,\alpha\d\alpha}=-2\epsilon_{\d\beta\d\alpha}
\Theta^{I_3}_\alpha,
\quad
 \Lambda^{I_3}_{\beta,\alpha\d\alpha}=-2\epsilon_{\beta\alpha}
\b\Theta^{I_3}_{\d\alpha}.
\end{equation}
Here, $\Theta^{I_3}_\alpha$ and $\b\Theta^{I_3}_{\d\alpha}$
are chiral and anti-chiral superfields, respectively:
\begin{equation}
 \b\na_{\d\beta} \Theta^{I_3}_{\alpha}=0,
\quad
 \na_\beta \b\Theta^{I_3}_{\d\alpha}=0.
\end{equation}
The gauge transformation laws of the prepotential $X^{I_3}$
are determined by those of $C^{I_3}_{\gamma\d\beta a}$:
\begin{equation}
 \delta_T(
\Lambda^{I_1},\Lambda^{I_2},\Theta_{\ul\beta}^{I_3},\Theta^{I_4}) 
 X^{I_3}
=\df{1}{2i}(\na^\alpha \Theta_\alpha^{I_3}-\b\na_{\d\alpha}\b\Theta^{I_3\d\alpha})
+(q^{(3)}\cdot \Theta)^{I_3}.
\label{eq:3gtpp}
\end{equation}
We find that $X^{I_3}$ are also transformed by the remaining 
gauge parameters $\Theta^{I_4}$ due to the tensor hierarchy.

The WZ gauge conditions for the prepotentials $X^{I_3}$ 
can be imposed by the parameters $\Theta^{I_3}_\alpha$ as follows:
\begin{equation}
 X^{I_3}|=0,
\quad
\na_\alpha X^{I_3}|=0,
\quad
\b\na_{\d\alpha} X^{I_3}|=0.
\label{eq:3wz}
\end{equation}
Note that 
the WZ conditions in Eq.~\eqref{eq:3wz} are imposed under the WZ gauge conditions 
for the prepotentials of 4-form gauge fields in Eq.~\eqref{eq:4wz}.

\parag{The 2-form gauge fields}
We find the remaining gauge parameters which leave the 
gauge-fixing conditions invariant:
\begin{equation}
 \delta_T(
\Lambda^{I_1},\Lambda^{I_2},\Theta_{\ul\gamma}^{I_3},\Theta^{I_4})
B_{\ul\beta\,\ul\alpha}^{I_2}=0.
\label{eq:2gi}
\end{equation}
We find that such parameters are given by
\begin{equation}
 \Lambda^{I_2}_\alpha =i \na_\alpha \Theta^{I_2},
\quad
 \Lambda^{I_2}_{\d\alpha} =-i \b\na_{\d\alpha} \Theta^{I_2},
\quad
 \Lambda^{I_2}_{\alpha\d\alpha}
=\df{1}{2}[\na_\alpha,\b\na_{\d\alpha}] \Theta^{I_2},
\end{equation}
where $\Theta^{I_2}$ are real superfields.
Again, $\Theta_{\ul\alpha}^{I_3}$ do not affect
 the gauge-fixing conditions in 
Eq.~\eqref{eq:2gi} in the presence of the tensor hierarchy.
The gauge transformation laws of the prepotential
$\Sigma^{I_2}_{\ul\alpha}$ are given by
\begin{equation}
 \begin{split}
  \delta_T(
 \Lambda^{I_1},\Theta^{I_2},\Theta_{\ul\beta}^{I_3},\Theta^{I_4})
 \Sigma^{I_2}_\alpha
 &=-\df{1}{4} \b\na^2\na_\alpha \Theta^{I_2}
 +(q^{(2)}\cdot \Theta_\alpha)^{I_2},
\\
 \delta_T(
\Lambda^{I_1},\Theta^{I_2},\Theta_{\ul\beta}^{I_3},\Theta^{I_4})
\b\Sigma^{I_2}_{\d\alpha}
&=-\df{1}{4} \na^2\b\na_{\d\alpha} \Theta^{I_2}
+(q^{(2)}\cdot \b\Theta_{\d\alpha})^{I_2}.
\end{split}
\label{eq:2gtpp}
 \end{equation}
Under the conditions in Eqs.~\eqref{eq:4wz} and \eqref{eq:3wz},
 we can go to the WZ gauge conditions for $\Sigma_\alpha^{I_2}$:
\begin{equation}
 \Sigma^{I_2}_\alpha|=0,
\quad
\b\Sigma^{I_2}_{\d\alpha}|=0,
\quad
(\na^\alpha \Sigma^{I_2}_\alpha
+\b\na_{\d\alpha}\b\Sigma^{I_2\d\alpha})|=0.
\label{eq:2wz}
\end{equation}
\parag{The 1-form gauge fields}
The gauge transformations for the 1-form gauge fields 
are the same as in ordinary super QED case except 
the shifts due to the tensor hierarchy.
We find that the gauge transformations which 
leave the gauge-fixing conditions 
in table \ref{tab:gf} invariant are given by 
\begin{equation}
 \Lambda^{I_1}= \df{1}{2}(\Theta^{I_1} + \b\Theta^{I_1}),
\end{equation}
Here, $\Theta^{I_1}$ and $\b\Theta^{I_1}$ 
are chiral and anti-chiral superfields, respectively:
\begin{equation}
 \b\na_{\d\alpha} \Theta^{I_1}=0,
\quad
 \na_\alpha \b\Theta^{I_1}=0.
\end{equation}
The gauge transformations of the 1-form prepotentials 
are given by the imaginary parts of $\Theta^{I_1}$
and the shifts by the gauge parameters of 2-form gauge fields
$\Theta^{I_2}$:
\begin{equation}
 \delta_T(
\Theta^{I_1},\Theta^{I_2},\Theta_{\ul\alpha}^{I_3},\Theta^{I_4})
 V^{I_1} 
= \df{1}{2i} (\Theta^{I_1}-\b\Theta^{I_1})
+(q^{(1)}\cdot \Theta)^{I_1}.
\label{eq:1gtpp}
\end{equation}
The WZ gauge conditions for the prepotentials $V^{I_1}$ can be 
imposed under the conditions in Eqs.~\eqref{eq:4wz}, \eqref{eq:3wz} and
\eqref{eq:2wz}: 
\begin{equation}
 V^{I_1}|=0, 
\quad 
\na_\alpha V^{I_1}|=0,
\quad
\b\na_{\d\alpha} V^{I_1}|=0,
\quad
\na^2 V^{I_1}|=0,
\quad
\b\na^2 V^{I_1}|=0.
\end{equation}
\parag{The 0-form gauge fields}
The gauge transformation laws of the prepotentials of
0-form are given by the chiral shifts by 
the gauge parameters $\Theta^{I_1}$: 
\begin{equation}
 \delta_T(
\Theta^{I_1},\Theta^{I_2},\Theta_{\ul\alpha}^{I_3},\Theta^{I_4})
 \Phi^{I_0}
=(q^{(0)}\cdot \Theta)^{I_0}.
\label{eq:0gtpp}
\end{equation}
Again, the shifts come from the tensor hierarchy.
\section{Chern--Simons actions}
\label{sec:cs}
In this section, we construct CS actions
in the conformal superspace.
The CS actions of the tensor hierarchy 
is related to anomaly cancellations in low energy 
effective theories.
The construction of the CS actions in the conformal superspace 
are quite similar to the global SUSY case \cite{bib:BBLR}.
CS actions are constructed by the combinations of the 
prepotentials and irreducible superfields 
$(Y^{I_3},L^{I_2},W_{\ul\alpha}^{I_1},\Psi^{I_0},J^{I_{-1}})$.

To construct the CS actions, we use the descent formalism.
This formalism systematically gives the CS actions 
from the internal transformation laws of the prepotentials.
We show that the descent formalism
 that was given in Ref.~\cite{bib:BBLR}
is straightforwardly extended in the case of 
the conformal superspace.
\parag{Descent formalism in global SUSY}
We briefly review the descent formalism in global SUSY in 
Ref.~\cite{bib:BBLR}.
The descent formalism in global SUSY is given by the 
combinations of the prepotentials and irreducible field strengths
as
\newcommand{\CS}{\text{CS}}
\newcommand{\re}{\text{Re}}
\begin{equation}
 S_{\CS}
=
\int d^4 x d^4\theta (V^{I_1}c_{I_1}-X^{I_3}c_{I_3})
+
\re\(
i\int d^4x d^2\theta 
(\Phi^{I_0}c_{I_0}+\Sigma^{I_2\alpha}c_{I_2\alpha}+\Gamma^{I_4}c_{I_4})
\).
\end{equation}
Here,  $c$'s are polynomials of the irreducible superfields 
$Y^{I_3}$, $L^{I_2}$, $W_{\ul\alpha}^{I_1}$,
$\Psi^{I_0}$ and $J^{I_{-1}}$.
The superfields $c_{I_1}$ and $c_{I_3}$ are 
real superfields, and 
$c_{I_0}$, $c_{I_2\alpha}$, and $c_{I_4}$ are 
chiral superfields.
The internal gauge invariance requires that 
$c$'s are related each other as
\begin{equation}
 \begin{split}
-\df{1}{4}\b{D}^2 c_{I_1}&=(q^{(0)})^{I_0}_{I_1}c_{ I_0},
\\
\df{1}{2i}
\(
D^\alpha c_{I_2 \alpha}
-\b{D}_{\d\alpha}\b{c}^{\d\alpha}_{ I_2}
\)
&=-(q^{(1)})^{I_1}_{I_2} c_{I_1},
\\
-\df{1}{4}\b{D}^2D_\alpha c_{ I_3}
&=(q^{(2)})^{I_2}_{I_3}c_{I_2 \alpha},
\\
\df{1}{2i}(c_{I_4}-\b{c}_{I_4})
&=-(q^{(3)})^{I_3}_{I_4}c_{I_3}.
 \end{split}
\end{equation}
Here, the derivatives $D_\alpha$ and $\b{D}_{\d\alpha}$ are
the covariant spinor derivatives in global SUSY:
$D_\alpha=\der_\alpha +i(\sigma^a)_{\alpha\d\alpha}
\b\theta^{\d\alpha} \der_a$ and
$\b{D}_{\d\alpha}=-\b\der_{\d\alpha} -i\theta^\alpha(\sigma^a)_{\alpha\d\alpha}
 \der_a$.
The internal gauge invariances are obtained by 
the relation between the superspace integrations:
\begin{equation}
 \int d^4x d^4\theta V=
-\df{1}{4}\int d^4x d^2\theta \b{D}^2 V
=
-\df{1}{4}\int d^4x d^2\b\theta D^2 V,
\label{eq:DtoFGS}
\end{equation}
where $V$ is a real superfield.
\parag{Descent formalism in the conformal superspace}
We now discuss the descent formalism in the conformal superspace.
The descent formalism in the conformal superspace 
is given by a natural extension of global SUSY case as
\begin{equation}
 S_{\CS}
=
\int d^4 x d^4\theta E(V^{I_1}c_{I_1}-X^{I_3}c_{I_3})
+
\re\(
i\int d^4x d^2\theta {\cal E}
(\Phi^{I_0}c_{I_0}+\Sigma^{I_2\alpha}c_{I_2\alpha}+\Gamma^{I_4}c_{I_4})
\),
\end{equation}
 where 
$E$ and ${\cal E}$ are the density of the whole superspace and
chiral subspace, respectively.
The integrations $\int d^4x d^4\theta E$ and 
$\int d^4x d^2\theta {\cal E}$ are called 
D- and F-term integration, respectively~\cite{bib:B1}.
The superfields 
$c$'s are polynomials of the irreducible superfields 
$Y^{I_3}$, $\b{Y}^{I_3}$, $L^{I_2}$, $W_{\ul\alpha}^{I_1}$,
$\Psi^{I_0}$ and $J^{I_{-1}}$.
Again, $c_{I_1}$ and $c_{I_3}$ are 
real superfields, and 
$c_{I_0}$, $c_{I_2\alpha}$, and $c_{I_4}$ are 
chiral superfields.
The $c$'s have two type of conditions.
One is the condition that is required by the 
superconformal invariance.
The conditions are that all the $c$'s are primary superfields,
and the conformal weights of them are as follows:
\begin{equation}
\begin{split}
 c_{I_0}&: (\Delta,w)=(3,2),
\\
 c_{I_1}&: (\Delta,w)=(2,0),
\\
 c_{I_2 \alpha}&: (\Delta,w)=(3/2,1),
\\
 c_{I_3}&: (\Delta,w)=(0,0),
\\
 c_{I_4}&: (\Delta,w)=(0,0).
\end{split}
\label{eq:cwc}
\end{equation}
The other is the condition that is required by  the internal gauge invariance of the tensor 
hierarchy as in the global SUSY case.
The internal gauge invariance requires 
the same conditions as those of Ref.~\cite{bib:BBLR}:
\begin{equation}
 \begin{split}
-\df{1}{4}\b\na^2 c_{I_1}&=(q^{(0)})^{I_0}_{I_1}c_{ I_0},
\\
\df{1}{2i}
\(
\na^\alpha c_{I_2 \alpha}
-\b\na_{\d\alpha}\b{c}^{\d\alpha}_{ I_2}
\)
&=-(q^{(1)})^{I_1}_{I_2} c_{I_1},
\\
-\df{1}{4}\b\na^2\na_\alpha c_{ I_3}
&=(q^{(2)})^{I_2}_{I_3}c_{I_2 \alpha},
\\
\df{1}{2i}(c_{I_4}-\b{c}_{I_4})
&=-(q^{(3)})^{I_3}_{I_4}c_{I_3}.
 \end{split}
\label{eq:desc}
\end{equation}
The internal gauge invariances are obtained by 
superspace partial integrations of the integrands.
In the conformal superspace, the relation between 
F-term and D-term actions is
\begin{equation}
 \int d^4x d^4\theta EV
=
-\df{1}{4}\int d^4x d^2\theta {\cal E}\b\na^2 V
=
-\df{1}{4}\int d^4x d^2\b\theta \b{\cal E}\na^2 V.
\label{eq:DtoF}
\end{equation}
Here, $V$ is a primary scalar superfield
with the conformal weight $(\Delta,w)=(2,0)$~\cite{bib:B1}.
Although the derivation of the relation between 
D- and F-term integrations is a bit nontrivial 
(see 
Ref.~\cite{bib:B1}), 
the relation is obtained by replacing
$d^4x d^4\theta$, $d^4x d^2\theta$, $D_{\alpha}$ and $\b{D}_{\d\alpha}$
in Eq.~\eqref{eq:DtoFGS}
with $ d^4x d^4\theta E$, $ d^4x d^2\theta {\cal E}$,
 $\na_\alpha$ and $\b\na_{\d\alpha}$, respectively.
This is a strong point of the conformal superspace approach:
The relations of the integrals 
are quite similar to the case of the global SUSY.

We can go to Poincar\'e SUGRA by imposing the superconformal 
gauge-fixing~\cite{{bib:B1},{bib:KU}}.
Because the CS actions are superconformally invariant without 
a compensator, the CS actions are not changed by 
the superconformal gauge-fixing conditions.

We finally show an example of the CS actions.
We consider an action which is a natural extension 
of the action proposed in Ref.~\cite{bib:BBLR}:
\begin{equation}
\begin{split}
 S_{\CS}
&:=\int d^4x d^4\theta E
(
\alpha_{I_1I_2} V^{I_1}L^{I_2}
-\alpha_{I_3 I_0}X^{I_3}\Psi^{I_0}
)\\
&\quad
+\re
\(i \int d^4x d^2 \theta {\cal E}
(\alpha_{I_0I_3} \Phi^{I_0}Y^{I_3} 
+\alpha_{I_2 I_1} \Sigma^{I_2\alpha }W_\alpha^{I_1}
+\alpha_{I_4I_{-1}} \Gamma^{I_4} J^{I_{-1}})
\). 
\end{split}
\label{eq:cs2}
\end{equation}
Here, $\alpha$'s are constant parameters.
This action is obtained by choosing $c$'s as follows:
\begin{equation}
 c_{I_0}
=\alpha_{I_0I_3} Y^{I_3},
\quad
c_{I_1}=\alpha_{I_1I_2} L^{I_2},
\quad
c_{I_2\alpha}=\alpha_{I_2 I_1}W^{I_1}_\alpha,
\quad
c_{I_3}=\alpha_{I_3I_0}\Psi^{I_0},
\quad
c_{I_4}=\alpha_{I_4I_{-1}}J^{I_{-1}}.
\end{equation}
This action satisfies the conformal weight conditions
in Eq.~\eqref{eq:cwc}
by using 
the conformal weights of the irreducible superfields in 
Eq.~\eqref{eq:cwsf} and those of $\Phi^{I_0}$ (for $J^{I_{-1}}$).
The internal invariance in Eq.~\eqref{eq:desc} 
requires the same conditions 
as the case of global SUSY~\cite{bib:BBLR}:
\begin{equation}
\begin{split}
\alpha_{I_1 I_2} (q^{(2)})^{I_2}_{I_3}
&=-\alpha_{I_0 I_3} (q^{(0)})^{I_0}_{I_1},
\\
\alpha_{I_2 I_1} (q^{(1)})^{I_1}_{I_2}
&=\alpha_{I_1 I_2} (q^{(1)})^{I_1}_{I_2}, 
\\
\alpha_{I_2 I_1} (q^{(2)})^{I_2}_{I_3}
 &=-\alpha_{I_3 I_0} (q^{(0)})^{I_0}_{I_1}, \\
\alpha_{I_4 I_{-1}} (q^{(-1)})^{I_{-1}}_{I_0}
&=\alpha_{I_3 I_0} (q^{(3)})^{I_3}_{I_4}.
\end{split} 
\end{equation}
\section{Conclusion}
\label{sec:conc}
In this paper, we have constructed the CS actions
of Abelian tensor hierarchy
in 4D ${\cal N}=1$ conformal superspace.
In section \ref{sec:pp}, 
the constraints on the field strengths have been  
solved in terms of the prepotentials with the
gauge-fixing conditions. 
The explicit forms are given in 
Eqs.~\eqref{eq:4pp}, 
 \eqref{eq:2pp}, \eqref{eq:0pp} and table \ref{tab:gf}. 
The conformal weights have been also determined 
 by the conformal weights of the vielbein.
We have obtained the relations between 
the prepotentials and irreducible superfields in 
table~\ref{tab:fspp}.
We have also obtained
the gauge transformation laws of the prepotentials in 
Eqs.~\eqref{eq:4gtpp}, \eqref{eq:3gtpp}, \eqref{eq:2gtpp},
 \eqref{eq:1gtpp} and \eqref{eq:0gtpp}.
The CS actions have been constructed 
in the conformal superspace
by using prepotentials in section \ref{sec:cs}.
The conformal weights of the $c$'s are determined in 
Eq.~\eqref{eq:cwc}.
We have shown that the descent formalism is 
mostly the same as the case of global SUSY as in 
Eq.~\eqref{eq:desc}.
Finally, the examples of CS couplings are exhibited
in Eq.~\eqref{eq:cs2}.
These examples are natual extensions of global SUSY case.

The CS actions in 4D ${\cal N}=1$ SUGRA, in particular
the action in Eq.~\eqref{eq:cs2},
 would be useful to discuss phenomenology such as 
inflation of the early universe~\cite{
{Kaloper:2008fb},{Kaloper:2011jz}}.
It would be interesting to 
embed the approach which was proposed in 
Ref.~\cite{Randall:2016zpo} into the conformal superspace.
\subsection*{Acknowledgements}
The author thanks Shuntaro Aoki, 
Tetsutaro Higaki and Yusuke Yamada
for useful discussions.
This work is supported by Research Fellowships of Japan Society
 for the Promotion of Science for Young Scientists Grant Number
16J03226.


\begin{thebibliography}{99}
\bibitem{Siegel:1979ai} 
  W.~Siegel,
  ``Gauge Spinor Superfield as a Scalar Multiplet,''
  Phys.\ Lett.\ B {\bf 85} (1979) 333.
\bibitem{bib:G} 
  S.~J.~Gates, Jr.,
  ``Super P Form Gauge Superfields,''
  Nucl.\ Phys.\ B {\bf 184} (1981) 381.
\bibitem{bib:SS}
  S.~J.~Gates, M.~T.~Grisaru, M.~Ro\v{c}ek and W.~Siegel,
  ``Superspace Or One Thousand and One Lessons in Supersymmetry,''
  Front.\ Phys.\  {\bf 58} (1983) 1
  [hep-th/0108200].
\bibitem{Muller:1985vga}
  M.~Muller,
  ``Supergravity in U(1) Superspace With a Two Form Gauge Potential,''
  Nucl.\ Phys.\ B {\bf 264} (1986) 292.
\bibitem{Cecotti:1987qr} 
  S.~Cecotti, S.~Ferrara and L.~Girardello,
  ``Massive Vector Multiplets From Superstrings,''
  Nucl.\ Phys.\ B {\bf 294} (1987) 537.
\bibitem{Binetruy:1996xw} 
  P.~Binetruy, F.~Pillon, G.~Girardi and R.~Grimm,
  ``The Three form multiplet in supergravity,''
  Nucl.\ Phys.\ B {\bf 477} (1996) 175
  [hep-th/9603181].
\bibitem{Ovrut:1997ur} 
  B.~A.~Ovrut and D.~Waldram,
  ``Membranes and three form supergravity,''
  Nucl.\ Phys.\ B {\bf 506} (1997) 236
  [hep-th/9704045].
\bibitem{Louis:2004xi}
  J.~Louis and W.~Schulgin,
  ``Massive tensor multiplets in N=1 supersymmetry,''
  Fortsch.\ Phys.\  {\bf 53} (2005) 235
  [hep-th/0410149].
\bibitem{D'Auria:2004sy}
  R.~D'Auria and S.~Ferrara,
  ``Dyonic masses from conformal field strengths in D even dimensions,''
  Phys.\ Lett.\ B {\bf 606} (2005) 211
  [hep-th/0410051].
\bibitem{Kuzenko:2004tn}
  S.~M.~Kuzenko,
  ``On massive tensor multiplets,''
  JHEP {\bf 0501} (2005) 041
  [hep-th/0412190].
\bibitem{Louis:2007nd}
  J.~Louis and J.~Swiebodzinski,
  ``Couplings of N=1 chiral spinor multiplets,''
  Eur.\ Phys.\ J.\ C {\bf 51} (2007) 731
  [hep-th/0702211 [HEP-TH]].
\bibitem{deWit:2005hv}
  B.~de Wit and H.~Samtleben,
  ``Gauged maximal supergravities and hierarchies of nonAbelian vector-tensor systems,''
  Fortsch.\ Phys.\  {\bf 53} (2005) 442
  [hep-th/0501243].
\bibitem{deWit:2008ta}
  B.~de Wit, H.~Nicolai and H.~Samtleben,
  ``Gauged Supergravities, Tensor Hierarchies, and M-Theory,''
  JHEP {\bf 0802} (2008) 044
  [arXiv:0801.1294 [hep-th]].
\bibitem{deWit:2008gc}
  B.~de Wit and H.~Samtleben,
  ``The End of the p-form hierarchy,''
  JHEP {\bf 0808} (2008) 015
  [arXiv:0805.4767 [hep-th]].
\bibitem{Hartong:2009az}
  J.~Hartong, M.~Hubscher and T.~Ortin,
  ``The Supersymmetric tensor hierarchy of N=1,d=4 supergravity,''
  JHEP {\bf 0906} (2009) 090
  [arXiv:0903.0509 [hep-th]].
\bibitem{bib:BBLR}
K.~Becker, M.~Becker, W.~D.~Linch and D.~Robbins,
``Abelian tensor hierarchy in 4D, N = 1 superspace,''  
JHEP {\bf 1603} (2016) 052
  [arXiv:1601.03066 [hep-th]].

\bibitem{Becker:2016rku}
  K.~Becker, M.~Becker, W.~D.~Linch and D.~Robbins,
  ``Chern-Simons actions and their gaugings in 4D, $N =$ 1 superspace,''
  JHEP {\bf 1606} (2016) 097
  [arXiv:1603.07362 [hep-th]].

\bibitem{Girardi:1986zn}
  G.~Girardi and R.~Grimm,
  ``Chern-simons Forms and Four-dimensional $N=1$ Superspace Geometry,''
  Nucl.\ Phys.\ B {\bf 292} (1987) 181.
\bibitem{LopesCardoso:1991ifk}
  G.~Lopes Cardoso and B.~A.~Ovrut,
  ``A Green-Schwarz mechanism for D = 4, N=1 supergravity anomalies,''
  Nucl.\ Phys.\ B {\bf 369} (1992) 351.
\bibitem{Adamietz:1992dk}
  P.~Adamietz, P.~Binetruy, G.~Girardi and R.~Grimm,
  ``Supergravity and matter: Linear multiplet couplings and Kahler anomaly cancellation,''
  Nucl.\ Phys.\ B {\bf 401} (1993) 257.
\bibitem{bib:B1}
  D.~Butter,
  ``N=1 Conformal Superspace in Four Dimensions,''
  Annals Phys.\  {\bf 325} (2010) 1026
  [arXiv:0906.4399 [hep-th]].
\bibitem{bib:KTVN}
  M.~Kaku, P.~K.~Townsend and P.~van Nieuwenhuizen,
``Properties of Conformal Supergravity,''
  Phys.\ Rev.\ D {\bf 17} (1978) 3179.
\bibitem{bib:KT}
  M.~Kaku and P.~K.~Townsend,
  ``Poincare Supergravity As Broken Superconformal Gravity,''
  Phys.\ Lett.\ B {\bf 76} (1978) 54.
\bibitem{bib:TVN}
  P.~K.~Townsend and P.~van Nieuwenhuizen,
  ``Simplifications of Conformal Supergravity,''
  Phys.\ Rev.\ D {\bf 19} (1979) 3166.
\bibitem{bib:FGVN}
  S.~Ferrara, M.~T.~Grisaru and P.~van Nieuwenhuizen,
  ``Poincare and Conformal Supergravity Models With Closed Algebras,''
  Nucl.\ Phys.\ B {\bf 138} (1978) 430.
\bibitem{bib:CFGVP}
  E.~Cremmer, S.~Ferrara, L.~Girardello and A.~Van Proeyen,
  ``Yang-Mills Theories with Local Supersymmetry: Lagrangian, Transformation Laws and SuperHiggs Effect,''
  Nucl.\ Phys.\ B {\bf 212} (1983) 413.
\bibitem{bib:KUigc} 
  T.~Kugo and S.~Uehara,
  ``Improved Superconformal Gauge Conditions in the $N=1$ Supergravity {Yang-Mills} Matter System,''
  Nucl.\ Phys.\ B {\bf 222} (1983) 125.
\bibitem{Kugo:1982cu}
  T.~Kugo and S.~Uehara,
  ``Conformal and Poincare Tensor Calculi in $N=1$ Supergravity,''
  Nucl.\ Phys.\ B {\bf 226} (1983) 49.

\bibitem{bib:KU} 
  T.~Kugo and S.~Uehara,
  ``$N=1$ Superconformal Tensor Calculus: Multiplets With External Lorentz Indices and Spinor Derivative Operators,''
  Prog.\ Theor.\ Phys.\  {\bf 73} (1985) 235.
\bibitem{bib:KKLVP}
  R.~Kallosh, L.~Kofman, A.~D.~Linde and A.~Van Proeyen,
  ``Superconformal symmetry, supergravity and cosmology,''
  Class.\ Quant.\ Grav.\  {\bf 17} (2000) 4269
   [Class.\ Quant.\ Grav.\  {\bf 21} (2004) 5017]
  [hep-th/0006179].
\bibitem{bib:WB}
  J.~Wess and J.~Bagger,
  ``Supersymmetry and supergravity,''
  Princeton, USA: Univ. Pr. (1992) 259 p

\bibitem{bib:BG2}
  P.~Bin\'etruy, G.~Girardi and R.~Grimm,
  ``Supergravity couplings: A Geometric formulation,''
  Phys.\ Rept.\  {\bf 343} (2001) 255
  [hep-th/0005225].
\bibitem{bib:KYY}
  T.~Kugo, R.~Yokokura and K.~Yoshioka,
  ``Component versus Superspace Approaches to D=4, N=1 Conformal
    Supergravity,''
PTEP {\bf 2016} (2016) no.7, 073B07
  [arXiv:1602.04441 [hep-th]].
\bibitem{Kugo:2016lum}
  T.~Kugo, R.~Yokokura and K.~Yoshioka,
  ``Superspace Gauge Fixing in Yang-Mills Matter Coupled Conformal Supergravity,''
  arXiv:1606.06515 [hep-th].
\bibitem{bib:AHYY}
  S.~Aoki, T.~Higaki, Y.~Yamada and R.~Yokokura,
  ``Abelian tensor hierarchy in 4D ${\cal N}=1$ conformal supergravity,''
  arXiv:1606.04448 [hep-th].

\bibitem{Kaloper:2008fb}
  N.~Kaloper and L.~Sorbo,
  ``A Natural Framework for Chaotic Inflation,''
  Phys.\ Rev.\ Lett.\  {\bf 102} (2009) 121301
  [arXiv:0811.1989 [hep-th]].
\bibitem{Kaloper:2011jz} 
  N.~Kaloper, A.~Lawrence and L.~Sorbo,
  ``An Ignoble Approach to Large Field Inflation,''
  JCAP {\bf 1103} (2011) 023
  [arXiv:1101.0026 [hep-th]].

\bibitem{Randall:2016zpo}
  S.~Randall,
  ``Supersymmetric Tensor Hierarchies from Superspace Cohomology,''
  arXiv:1607.01402 [hep-th].
\end{thebibliography}
\end{document}